\documentclass{Interspeech2024}

\usepackage{comment, nccmath} 
\usepackage{multirow}
\usepackage[table,xcdraw]{xcolor}
\usepackage{cite,balance}

\interspeechcameraready

\title{How Do Neural Spoofing Countermeasures Detect Partially Spoofed Audio?}
\name[affiliation={1,2}]{Tianchi}{Liu}
\name[affiliation={3}]{Lin}{Zhang}
\name[affiliation={4}]{Rohan Kumar}{Das}
\name[affiliation={2}]{Yi}{Ma}
\name[affiliation={2}]{Ruijie}{Tao}
\name[affiliation={5,2}]{Haizhou}{Li}

\address{
\normalsize
  $^1$Institute for Infocomm Research (I$^2$R), Agency for Science, Technology and Research (A$^\star$STAR), Singapore\\
  $^2$Department of Electrical and Computer Engineering, National University of Singapore, Singapore \\
  $^3$National Institute of Informatics, Tokyo, Japan~~~  $^4$Fortemedia Singapore, Singapore\\
  $^5$School of Data Science, Shenzhen Research Institute of Big Data, The Chinese University of Hong Kong, Shenzhen, China} 

\email{ {\scriptsize \{tianchi\underline{\enskip}liu, mayi, ruijie.tao\}@u.nus.edu, zhanglin@nii.ac.jp, rohankd@fortemedia.com, haizhouli@cuhk.edu.cn} }


\begin{document}

\maketitle

\begin{abstract}
    

Partially manipulating a sentence can greatly change its meaning. Recent work shows that countermeasures (CMs) trained on partially spoofed audio can effectively detect such spoofing. However, the current understanding of the decision-making process of CMs is limited. We utilize Grad-CAM and introduce a quantitative analysis metric to interpret CMs' decisions. We find that CMs prioritize the artifacts of transition regions created when concatenating bona fide and spoofed audio. This focus differs from that of CMs trained on fully spoofed audio, which concentrate on the pattern differences between bona fide and spoofed parts. Our further investigation explains the varying nature of  CMs' focus while making correct or incorrect predictions. These insights provide a basis for the design of CM models and the creation of datasets. Moreover, this work lays a foundation of interpretability in the field of partial spoofed audio detection that has not been well explored previously. 

\end{abstract}

\section{Introduction}
\label{sec_intro}

The advancement of generative models in speech processing has magnified the threat of spoofed speech. To mitigate the potential damage from the synthetic speech, several advanced CMs have been proposed \cite{WU2015130, bib:Attacker_overview2020, wangxininterspeech, tak22_odyssey, 10155166, 9417604, 9747766, kawa23b_interspeech, zang23_interspeech}. Most of them focus on the fully spoofed scenario, where the spoofed audio is entirely created by methods such as text-to-speech, voice conversion, and so on.

\begin{figure}[t]
\centerline{\hspace{-0.4cm}\includegraphics[scale=0.029]{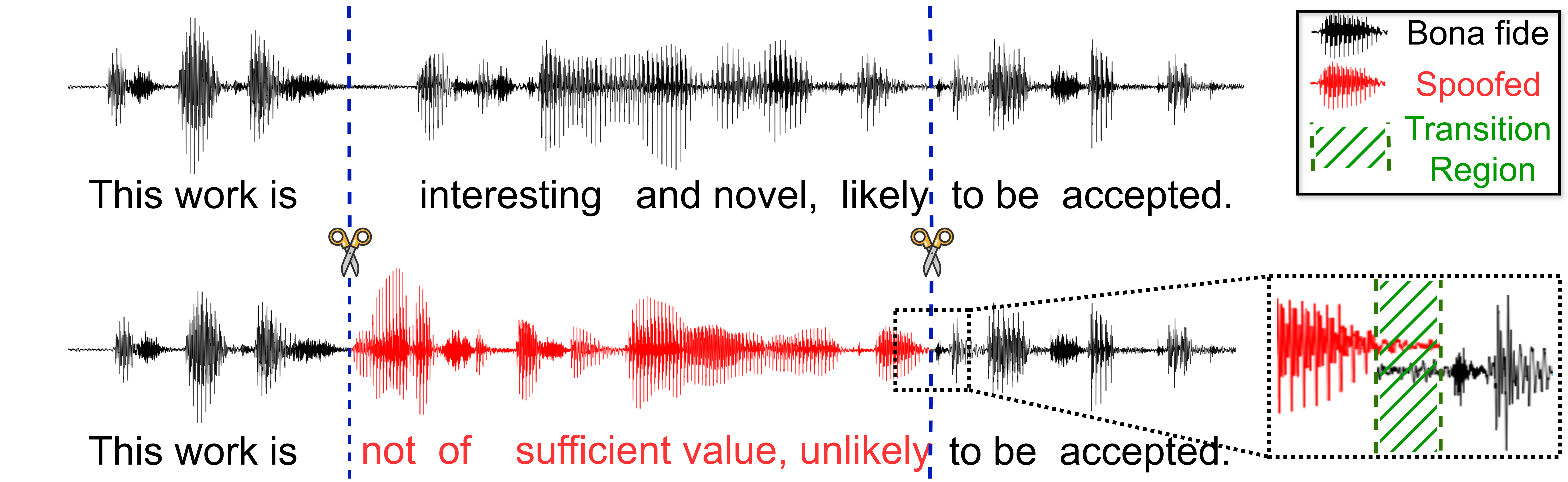}}
\vspace{-1mm}
\caption{Illustration of partially spoofed speech changing the meaning of a sentence.} 
\label{fig_example}
\vspace{-5mm}
\end{figure}

However, it is not always necessary to generate the entire audio with generative models from the perspective of spoofing. When attackers have access to bona fide samples close to the desired audio, they only need to manipulate arbitrary short parts using generative models. An example is depicted in Figure~\ref{fig_example}. This has led to a new spoofing scenario referred to as `Partial Spoof' (PS), which has attracted increased attention recently~\cite{zhanglin_PartialSpoof_TASLP, yi2023add, 10094774, muller2024mlaad}. In the PS scenario, when concatenating bona fide and spoofed segments, noticeable artifacts of discontinuity will occur. 
To address this, several strategies are applied, such as signal-processing techniques~\cite{zhanglin_PartialSpoof_TASLP, Yi2021halftruth}, neural network-based approaches \cite{cai2023avdf1m}, etc. These strategies aim to mitigate the discontinuity and create smoother transitions. Hence, we describe the resulting segments as transition regions (also known as `concatenated parts'~\cite{zhanglin_PartialSpoof_TASLP}), illustrated in Figure~\ref{fig_example}. 

Given that the spoofed region can be arbitrarily short, detecting spoofed audio in the PS scenario becomes more challenging, especially when training data consists entirely of fully spoofed samples \cite{Zhang2021PartialSpoofInitial}.
To improve the performance of detecting partially spoofed audio in the PS scenario, several neural CMs and data sets are proposed~\cite{zhanglin_PartialSpoof_TASLP, yi2023add, martin2022ADD4, 10094774, 9746162, muller2024mlaad, CAI2024101597, 9746939}. While these efforts remarkably improve the capability of detecting PS audio, the inherent characteristics of data-driven networks result in them functioning as \textit{black-boxes} with restricted interpretability. 

Understanding CMs is crucial not only for ensuring the trustworthiness of their decisions but also for benefiting the optimization of the model and the design of the training dataset. Therefore, it is essential to understand how CMs work. In light of this, there is a growing interest in the field of \textit{Explainable AI (XAI)} for CMs.
The authors of~\cite{silencegolden} explored the role of the silence region by analyzing its duration for a CM's effectiveness, while the work in~\cite{10224301}, analyzed the effect of removing silence and employed gradient-weighted class activation mapping (Grad-CAM) for visualization.
Along similar directions, the authors of~\cite{salvi2024listening} delved into the analysis of background noise in spoofed speech, whereas the study in~\cite{tak20_odyssey} focused on analyzing frequency bands and speech formants.
The Grad-CAM-Binary example is used in~\cite{halpern20_odyssey} to illustrate that buzziness and rhythm of speech are the determining factors for CMs. Similarly, SHapley Additive exPlanations (SHAP) contributes to exploring the artifacts produced by speech generators~\cite{9747476, ge22_odyssey}.
The existing studies mainly center on models trained with entirely spoofed data, but there is very limited exploration regarding partial spoofing CMs. This motivated us to add a new dimension of interpretability to the field of partially spoofed audio detection, an aspect that has not been thoroughly explored yet.

In this work, we aim to understand and explain how partial-spoofing CMs make decisions, thereby providing valuable insights for their development. We consider Grad-CAM \cite{gradcam} which uses the gradients to produce a localization map highlighting the important regions of the input for the model's predictions. While Grad-CAM has been applied in speech-related tasks~\cite{10224301, halpern20_odyssey, li2024phonemes, ma2024gradient}, to the best of our knowledge, this is the first effort to use it for explaining partial-spoofing CMs. We believe that compared to bona fide and spoofed speech parts, the model pays more attention to the artifacts from the transition regions between them. To validate this, we propose a novel metric to quantitatively analyze Grad-CAM scores. Additionally, we investigate the areas that the CM focuses on for the cases of successful and failed detection, offering valuable insights for the refinement of partial-spoofing CMs.

\section{Explaining CMs with Grad-CAM}

\subsection{Grad-CAM}

\begin{figure}[h]
\vspace{-0.15 in}
\centerline{\includegraphics[scale=0.052]{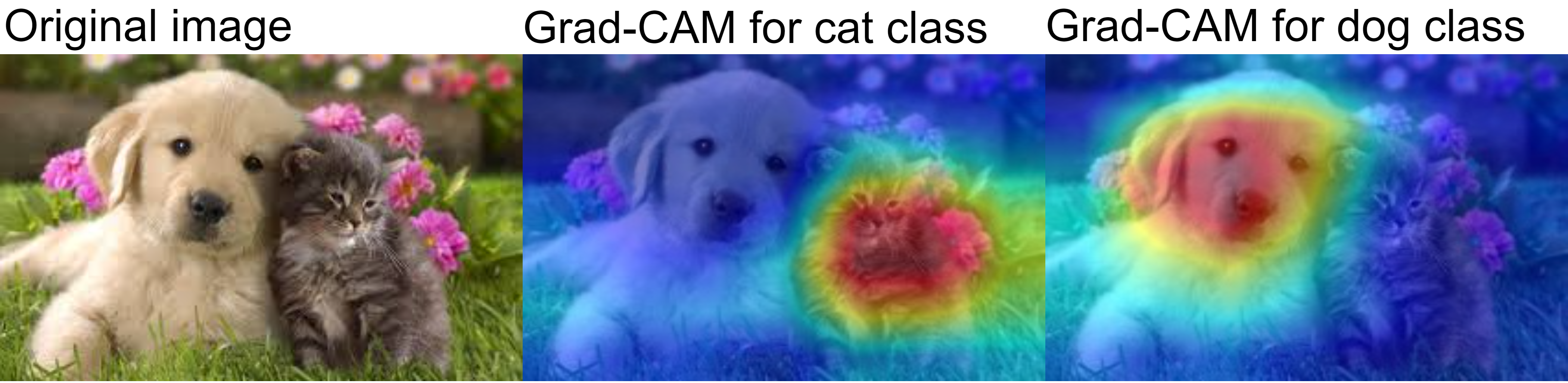}}
\vspace{-0.08 in}
\caption{An example of counterfactual explanations with Grad-CAM~\cite{gradcam}. Figures are produced using~\cite{jacobgilpytorchcam}.}
\label{fig_gradcamexample}
\vspace{-0.12 in}
\end{figure}

Grad-CAM is designed to utilize the gradients of a specified target output class, generating a coarse localization map that evaluates each input pixel's activation values. The larger values indicate that the corresponding regions are more important for the model’s predictions~\cite{gradcam}.

For an image, assuming $l_k=f(x)$ is the logit of the target class $k$  generated from the pretrained classification model, $R\in\mathbb{R}^{(C\times W\times H)}$ is the hidden representation from the target layer in dimension of \textit{channel} $\times$ \textit{width} $\times$ \textit{height}, the Grad-CAM score $S$ can be defined as follows:

\vspace{-0.10 in}
\begin{equation}
    S=ReLU(\sum_c R_{c,w,h} \cdot g_{c,w,h}),  \ g_{c,w,h}=\frac{\partial l_{k}}{\partial R_{c,w,h}}.
\end{equation}
\vspace{-0.15 in}

Specifically, the Grad-CAM is the multiplication of the hidden representation and the gradient from the target logits with respect to the hidden representation. The ReLU processing keeps the pixel having positive contributions to the prediction.

An example is illustrated in Figure~\ref{fig_gradcamexample}, where Grad-CAM identifies the crucial area in the image with large values of $S$. This highlights that the area contributes most significantly to the model's final decision, whether predicting a cat or a dog.

\subsection{Explaining CMs with Grad-CAM}
\label{subsec_explainCM}

\begin{figure}[t]
\vspace{-0.05 in}
\centerline{\includegraphics[scale=0.096]{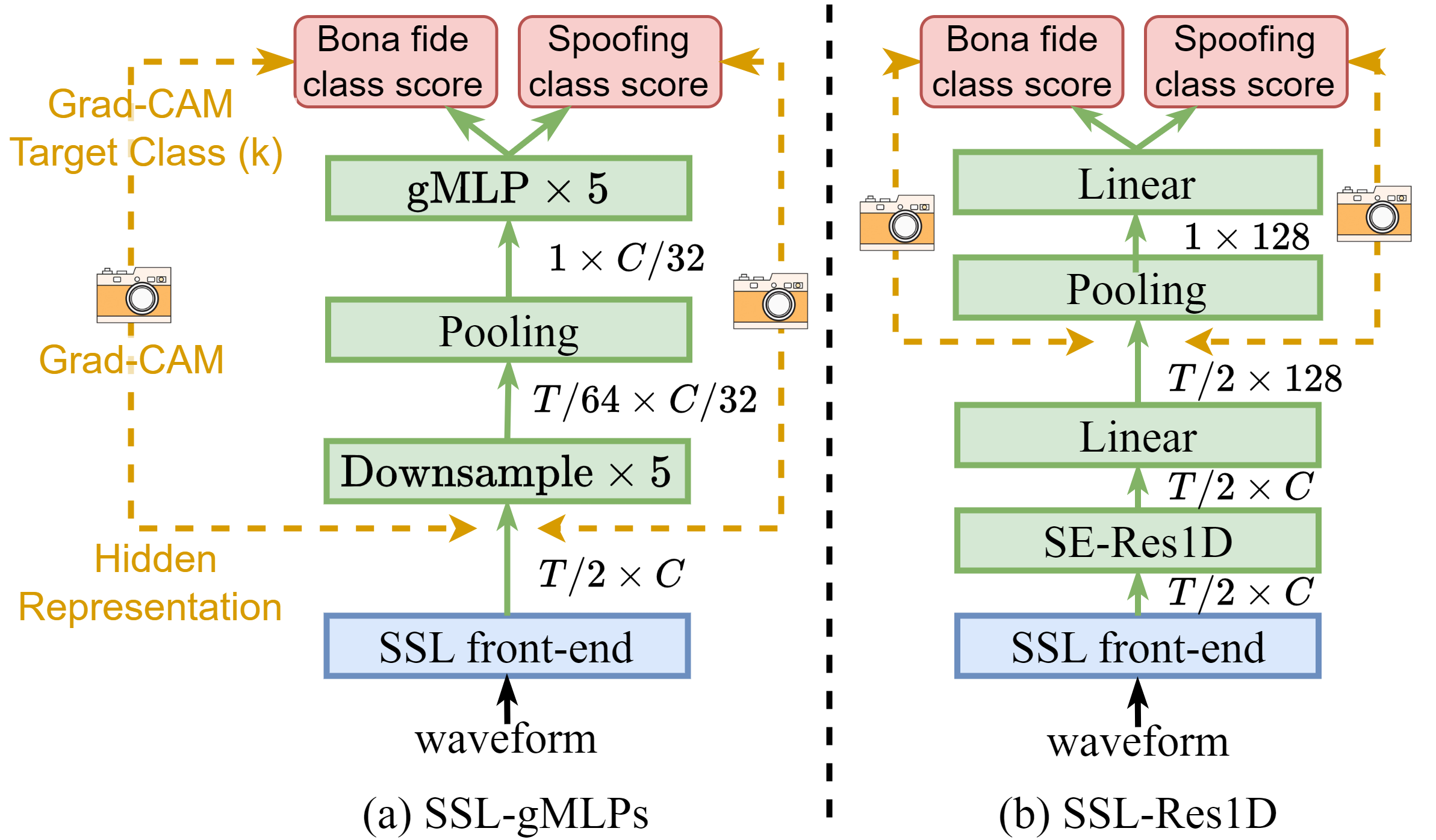}}
\vspace{-1.7 mm}
\caption{Block diagrams illustrating the structures of the (a) SSL-gMLPs and (b) SSL-Res1D models. 
The dashed yellow line with a camera icon indicates the Grad-CAM.}
\label{fig_model}
\vspace{-0.16 in}
\end{figure}

We adopt a recent top performing partial-spoofing CM in~\cite{zhanglin_PartialSpoof_TASLP} for analysis. It is structured with a self-supervised learning (SSL) model~\cite{NEURIPS2020_92d1e1eb} as front-end and stacked gated multilayer perceptron (gMLP)~\cite{NEURIPS2021_4cc05b35} layers as a classifier, shown in Figure~\ref{fig_model} (a). The downsampling block combines max pooling with a kernel size of $2$ and a linear layer. The output of SSL-gMLPs in~\cite{zhanglin_PartialSpoof_TASLP} adopts a two-class design of bona fide and spoofing.

To address the \textit{black-box} nature of partial spoofing CMs, we apply Grad-CAM to explain their decision-making process. This method helps analyze the regions that partial spoofing CMs focus on for their detection decisions. As shown in Figure~\ref{fig_model} (a), to obtain high-resolution Grad-CAM scores for accurate counterfactual explanations, we select the layer following the SSL front-end as the target layer for Grad-CAM. This choice is due to the model design described in~\cite{zhanglin_PartialSpoof_TASLP}, which employs max pooling during the downsampling stage, thereby reducing the time-domain resolution~\cite{10497864}.

\begin{figure*}[t]
\vspace{-0.06 in}
\centerline{\includegraphics[scale=0.083]{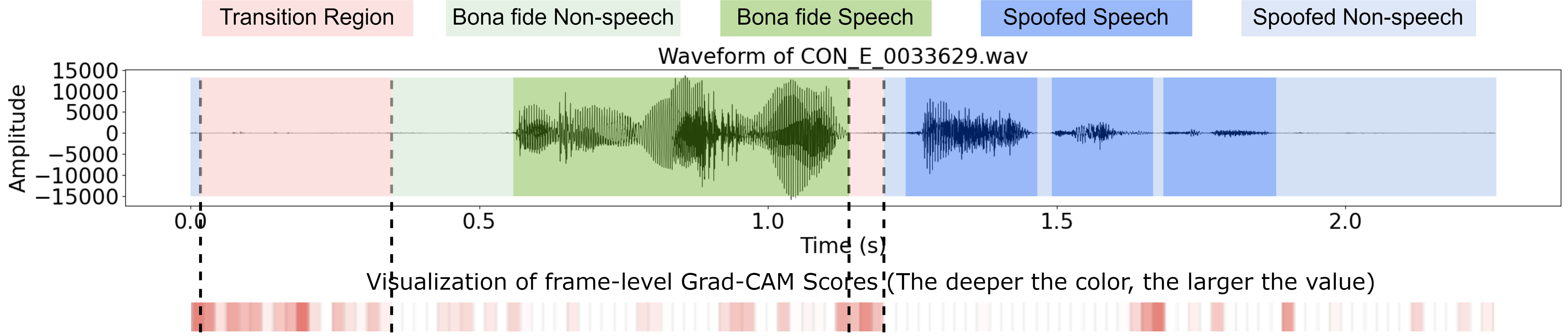}}
\vspace{-0.05 in}
\caption{Visualization of the waveform and frame-level Grad-CAM scores for \textit{CON\_E\_0033629.wav} from the evaluation set.}
\label{fig_visual_grad}
\vspace{-0.07 in}
\end{figure*}

 Considering the linear layers in multiple downsampling blocks and the stacking of multiple gMLPs, the class-discriminative information may become more diluted. This dilution could lead to less accurate or interpretable Grad-CAM scores due to the backward propagation through too many layers. Moreover, the layer immediately following the SSL front-end primarily captures basic speech representations, which may not meaningfully correlate with the CM's decision-making process in terms of explainability. To address these issues, we optimize the network, as shown in Figure~\ref{fig_model} (b), maintaining a high resolution until the last two layers of a pooling and a linear. The SE-Res1D is used to replace original gMLPs, combining a single 1D-ResNet block with squeeze-and-excitation (SE) module~\cite{ECAPA_TDNN, MFA_TDNN, liu2024disentangling}. The design of this new model aims to overcome the above-mentioned two challenges that affect explainability in the original SSL-gMLPs model. This effort leads to an improved version, which we refer to as SSL-Res1D.

Furthermore, in our effort to explain CMs' decision-making process through Grad-CAM, we aim to obtain accurate and high-resolution Grad-CAM activation values at the frame level for analysis. In this case, we refrain from using any score smoothing techniques or scaling up the frame-level activation scores to match the original waveform's shape via interpolation.

\vspace{-0.02 in}
\subsection{How to Analyze Grad-CAM Scores Quantitatively?}
\label{subsec_quant}
Visualizing through Grad-CAM scores, as shown in Figure~\ref{fig_gradcamexample}, is highly intuitive but falls short in quantitatively analyzing the entire dataset to draw a dependable conclusion.
To quantitatively validate our findings and provide a comprehensive numerical summary of the analysis across the entire dataset, we introduce a metric termed Relative Contribution Quantification (RCQ). The metric is designed to describe the relative relationships between different speech segments concerning their contribution to the final decision.

First, we use the Grad-CAM to derive the frame-level Grad-CAM scores of the $i$-th sample $\{S_{i,t=1}, S_{i,t=2}, \dots, S_{i,t=T}\}$ in the test set $I$, corresponding to the input speech sample of frame $T$. Then we align these scores with the frame-level annotations $y \in \{\text{TR, BS, BN, SS, SN} \}$ which represent five categories: \textbf{T}ansition \textbf{R}egion, \textbf{B}ona fide \textbf{S}peech, \textbf{B}ona fide \textbf{N}on-speech, \textbf{S}poofed \textbf{S}peech, and \textbf{S}poofed \textbf{N}on-speech, respectively. For each category, we compute the average Grad-CAM score $\bar{S}$ by averaging Grad-CAM scores across all $I$ samples in the test set. To illustrate, consider the category of transition region (TR):

\vspace{-0.08 in}
\begin{equation}
   \bar{S}_\text{TR} = \frac{1}{F_\text{TR}} \sum_{i=1}^{I}\sum_{t=1}^{T_{i}}S_{i,t}, \forall_{i,t} \ \text{s.t.} \ y_{i,t}=\text{TR},
\end{equation}
\vspace{-0.12 in}

\noindent where $F_\text{TR}$ is the total number of frames in the transition region category across the entire test set. In a similar manner, we compute the average Grad-CAM scores for the remaining categories, denoted as $\bar{S}_{\text{BS}}$, $\bar{S}_{\text{BN}}$, $\bar{S}_{\text{SS}}$, and $\bar{S}_{\text{SN}}$. 
Subsequently, we compute the RCQ for all five categories by evaluating their respective relative values to the average Grad-CAM score of all frames $\bar{S}_{\text{all}}$.
The calculation is exemplified for the transition region (TR) category as follows:

\vspace{-0.05 in}
\begin{equation}
   RCQ_\text{TR} =  (\bar{S}_\text{TR} - \bar{S}_{\text{all}}) \div \bar{S}_{\text{all}} \times 100\%.
   \label{eq_avg}
\end{equation}
\vspace{-0.19 in}

The RCQ metric quantitatively evaluates the contribution of each category to the final decision across the entire dataset.
It also provides insight into the relative importance of the five categories, {where a larger value indicates greater importance}.

\section{Experimental Setup}
\vspace{-0.05 in}
\subsection{Dataset}
In this work, we utilized the public accessible database, PartialSpoof~\cite{zhanglin_PartialSpoof_TASLP}, which is notable for providing detailed timestamp annotations. These annotations cover both speech and non-speech parts from bona fide and spoof, as well as transition regions. This comprehensive annotation enables its use in studies aimed at an in-depth understanding of how CMs make decisions.

\subsection{Configuration}
\vspace{-0.05 in}

We mostly follow the study in~\cite{zhanglin_PartialSpoof_TASLP} and its recipe\footnote{\scriptsize\url{https://github.com/nii-yamagishilab/PartialSpoof}} to develop CMs. Specifically, we use the wav2vec2-large~\cite{NEURIPS2020_92d1e1eb} as the front-end, \textit{MSE for P2SGrad} loss~\cite{wangxininterspeech} as the loss function, and Adam as optimizer. To accommodate GPU memory limitations, we reduce the batch size to $2$, and adjust the learning rate accordingly to $2.5 \times 10^{-6}$. Training is stopped if there is no improvement on the development set for $20$ epochs. The optimal epoch on the development set is selected to report the performance and be analyzed by Grad-CAM. The equal error rate (EER) metric is used to report the performance of detecting spoofed audio.

For Grad-CAM, we adopt the PyTorch library for CAM methods~\cite{jacobgilpytorchcam}\footnote{\scriptsize\url{https://github.com/jacobgil/pytorch-grad-cam}}, with necessary modification for transitioning from images to speech signals, which involves one dimension less. 
The frame length for Grad-CAM analysis is $20$ ms~\cite{zhanglin_PartialSpoof_TASLP}, constrained by the output dimension of wav2vec2-large~\cite{NEURIPS2020_92d1e1eb}.

\section{Results and Discussion}

\subsection{Partially Spoofed Speech Detection Evaluation}

Table~\ref{tab:ps} reports the performance for the SSL-gMLPs and SSL-Res1D models on the development and evaluation sets of PartialSpoof. A notable performance improvement is observed when utilizing the PartialSpoof dataset for training, compared to using the fully spoofed ASVspoof dataset. 
Moreover, the SSL-Res1D model shows superior performance in comparison to SSL-gMLPs when trained on the ASVspoof 2019 LA dataset and closely matches the top-tier results in~\cite{zhanglin_PartialSpoof_TASLP} upon training with the PartialSpoof dataset. These results illustrate that the models we analyzed are comparable to those in~\cite{zhanglin_PartialSpoof_TASLP}. 

\begin{table}[h]
\footnotesize
\centering
\caption{Performance in EER (\%) for SSL-gMLPs and SSL-Res1D, trained on the ASVspoof 2019 LA~\cite{ASVspoof2019data} or PartialSpoof~\cite{zhanglin_PartialSpoof_TASLP} dataset and tested on the development as well as evaluation sets of PartialSpoof. $\dag$ indicates re-implementation.}
\vspace{-0.05 in}
\setlength{\tabcolsep}{1.7mm}{
\begin{tabular}{lccc}
\hline
\toprule
Model & Training Set & Dev. Set & Eval. Set \\ \hline
\midrule
SSL-gMLPs~\cite{zhanglin_PartialSpoof_TASLP} &  ASVspoof 2019 LA & 12.22 & 14.19\\
SSL-gMLPs$\dag$ &  ASVspoof 2019 LA & 12.75 & 13.17 \\
\textbf{SSL-Res1D (ours) }&  ASVspoof 2019 LA & \textbf{4.67} & \textbf{3.60} \\
\hline
SSL-gMLPs~\cite{zhanglin_PartialSpoof_TASLP} &  PartialSpoof  & \textbf{0.35} & \textbf{0.64}\\
SSL-gMLPs$\dag$ &  PartialSpoof & 0.51 & 0.86 \\
\textbf{SSL-Res1D (ours) } &  PartialSpoof & \textbf{0.35} & 0.73 \\

\bottomrule
\hline
\end{tabular}
}
\label{tab:ps}
\vspace{-0.18 in}
\end{table}

\subsection{Visualizing Frame-level Grad-CAM}
A visualization example of the Grad-CAM score for each frame is shown in Figure~\ref{fig_visual_grad}. As introduced in Section~\ref{subsec_quant}, partially spoofed audio in the PartialSpoof database comprises five components: the non-speech as well as speech segments of both bona fide and spoofed audio, and overlap-add-based transition region. The Grad-CAM scores for each frame are displayed in the lower panel of Figure~\ref{fig_visual_grad}, where a deeper color shade indicates a higher value. These Grad-CAM values signify each frame's contribution to the final prediction. Observations from Figure~\ref{fig_visual_grad} reveal that the CM primarily focuses on the transition region considering the evaluation sample \textit{CON\_E\_0033629.wav}.

\begin{table*}[t]
\footnotesize
\centering
\vspace{-0.06 in}
\caption{
RCQs (\%) of SSL-Res1D models when predicting spoof and bona fide classes' scores for partially spoofed samples, across five different segment types. Models are trained on ASVspoof 2019 LA~\cite{ASVspoof2019data} or PartialSpoof~\cite{zhanglin_PartialSpoof_TASLP}.
The \colorbox{gray!30}{grey} color represents the relative size relationship among the values of the five types of segments within each trial, with deeper shades indicating larger values.
}
\vspace{-0.06 in}
\setlength{\tabcolsep}{0.5mm}{

\begin{tabular}{cccrrrrrcrrrrr}

\hline
\toprule
 &  & \multicolumn{6}{c}{\underline{\qquad \qquad \qquad \ PartialSpoof Development Set \qquad \qquad \qquad \ }} & \multicolumn{6}{c}{\underline{\qquad \qquad \qquad \ 
 PartialSpoof Evaluation Set \qquad \qquad \qquad \ }} \\
 &  &  & \multicolumn{1}{c}{Bona fide} & \multicolumn{1}{c}{Spoofed} & \multicolumn{1}{c}{Transition} & \multicolumn{1}{c}{Bona fide} & \multicolumn{1}{c}{Spoofed} &  & \multicolumn{1}{c}{Bona fide} & \multicolumn{1}{c}{Spoofed} & \multicolumn{1}{c}{Transition} & \multicolumn{1}{c}{Bona fide} & \multicolumn{1}{c}{Spoofed} \\
\multirow{-3}{*}{Training Set} & \multirow{-3}{*}{\begin{tabular}[c]{@{}c@{}}Grad-CAM \\ Target\\ Class (k)\end{tabular}} & \multirow{-2}{*}{\begin{tabular}[c]{@{}c@{}}Utt.\\ EER\end{tabular}} & \multicolumn{1}{c}{Speech} & \multicolumn{1}{c}{Speech} & \multicolumn{1}{c}{{Region}} & \multicolumn{1}{c}{Non-speech} & \multicolumn{1}{c}{Non-speech} & \multirow{-2}{*}{\begin{tabular}[c]{@{}c@{}}Utt.\\ EER\end{tabular}} & \multicolumn{1}{c}{Speech} & \multicolumn{1}{c}{Speech} & \multicolumn{1}{c}{Region} & \multicolumn{1}{c}{Non-speech} & \multicolumn{1}{c}{Non-speech} \\

 \hline
\midrule

 & Spoof &  & \cellcolor[HTML]{D4D4D4}-3.60 & \cellcolor[HTML]{999999}31.38 & \cellcolor[HTML]{C7C7C7}4.07 & \cellcolor[HTML]{FFFFFF}-29.77 & \cellcolor[HTML]{E4E4E4}-13.42 &  & \cellcolor[HTML]{CCCCCC}-3.62 & \cellcolor[HTML]{999999}25.81 & \cellcolor[HTML]{A1A1A1}21.69 & \cellcolor[HTML]{FFFFFF}-33.43 & \cellcolor[HTML]{D3D3D3}-7.87 \\
\multirow{-2}{*}{ASVspoof19} & Bona fide & \multirow{-2}{*}{4.67} & \cellcolor[HTML]{D0D0D0}-0.77 & \cellcolor[HTML]{FFFFFF}-35.07 & \cellcolor[HTML]{999999}39.14 & \cellcolor[HTML]{A6A6A6}30.24 & \cellcolor[HTML]{BEBEBE}12.75 & \multirow{-2}{*}{3.60} & \cellcolor[HTML]{E0E0E0}-7.11 & \cellcolor[HTML]{FFFFFF}-27.24 & \cellcolor[HTML]{9C9C9C}37.03 & \cellcolor[HTML]{999999}38.72 & \cellcolor[HTML]{BEBEBE}15.26 \\
\hline
 & Spoof &  & \cellcolor[HTML]{E0E0E0}-0.06 & \cellcolor[HTML]{D7D7D7}1.87 & \cellcolor[HTML]{999999}15.03 & \cellcolor[HTML]{FFFFFF}-6.95 & \cellcolor[HTML]{E8E8E8}-1.81 &  & \cellcolor[HTML]{F2F2F2}-0.62 & \cellcolor[HTML]{F3F3F3}-0.90 & \cellcolor[HTML]{999999}25.36 & \cellcolor[HTML]{FFFFFF}-4.72 & \cellcolor[HTML]{ECECEC}0.99 \\
\multirow{-2}{*}{PartialSpoof} & Bona fide & \multirow{-2}{*}{0.35} & \cellcolor[HTML]{F8F8F8}-3.65 & \cellcolor[HTML]{FFFFFF}-7.90 & \cellcolor[HTML]{999999}49.85 & \cellcolor[HTML]{EFEFEF}1.27 & \cellcolor[HTML]{F0F0F0}0.89 & \multirow{-2}{*}{0.73} & \cellcolor[HTML]{FFFFFF}-8.63 & \cellcolor[HTML]{F9F9F9}-4.57 & \cellcolor[HTML]{999999}58.01 & \cellcolor[HTML]{E3E3E3}10.22 & \cellcolor[HTML]{F4F4F4}-1.09 \\

\bottomrule
\hline
\end{tabular}
}
\label{tab:quanti}
\vspace{-0.142 in}
\end{table*}

\subsection{Quantitative Analysis of Grad-CAM Scores}
\label{subsec_quantianaly}
The visualization is intuitive but limited to one example. For quantitative analysis, we apply the RCQ metric from Section~\ref{subsec_quant}. Our focus is on partially spoofed samples, as bona fide samples in PartialSpoof contain only bona fide parts, excluding spoofed parts and transition regions.

Figure~\ref{fig_model} illustrates that each model generates two scores for a sample.
These two scores are considered as the probabilities that the utterance is bona fide and spoofed, respectively. Ideally, for a spoofed utterance, the score for the spoof class would be large, whereas the score for the bona fide class would be small. We employ Grad-CAM to separately compute the gradients for these two target classes ($k$), obtaining  the Grad-CAM scores $S$ and deriving RCQs. In this case, we can analyze different focuses of the CMs when predicting scores for these two target classes. The RCQs are reported in Table~\ref{tab:quanti}. Below are our discussions and findings:

\textbf{Differences in focus between CMs trained on fully vs. partially spoofed data:} Models trained with ASVspoof and PartialSpoof data show different focuses when detecting partially spoofed audio. 
The ASVspoof-trained model primarily focuses on spoofed speech when predicting scores for the spoof output class. Conversely, it gives greater emphasis to non-speech, especially bona fide non-speech segments for scoring the bona fide output class. 
The focus on transition regions, is likely because these regions are predominantly non-speech. In contrast, models trained with PartialSpoof data seem to find an effective solution by focusing mainly on the transition regions.

\textbf{What is important for CM to learn in PS?} The CM trained on the PartialSpoof dataset effectively recognized the transition regions' pattern, a feature not encountered by the CM trained on the ASVspoof2019 dataset.
Linking the difference in learned focus to the EER, and considering the identical structure and training strategy, we believe that focusing on transition regions is remarkably effective and efficient when dealing with the PartialSpoof dataset.

\textbf{Varied focus of CMs in predicting scores for spoof and bona fide classes: } Both the ASVspoof2019 and PartialSpoof trained models show noticeably different focuses when predicting spoof and bona fide class scores. The ASVspoof2019 trained model pays more attention to spoofed speech for its spoof class, while prioritizing non-speech for its bona fide class, especially bona fide non-speech. This suggests that the model not only learns the differences in pattern between bona fide and spoofed segments, but also autonomously discerns that speech and non-speech information are respectively suited for different class predictions: non-speech features are more effective for bona fide class predictions, while characteristics of spoofed speech are important for spoof class predictions.
Moreover, the ASVspoof2019 trained model strongly emphasizes non-speech over speech when predicting bona fide class, providing insights and evidence into the importance of the non-speech aspect following previous studies~\cite{silencegolden, 10224301}. This also highlights the value of this explanatory work. Additionally, the CM trained on the PartialSpoof dataset mostly focuses on the transition regions, this applies to predicting scores for both spoof and bona fide classes. It also exhibits a subtly increased attention to the segments that are particularly relevant to each class.

\subsection{What Do CMs Focus on Successful vs. Failed Cases?}

\begin{table}[h]
\footnotesize
\centering
\vspace{-0.08 in}
\caption{RCQs (\%) of five different types of speech segments for misclassified partially spoofed samples with SSL-Res1D model.}
\vspace{-0.06 in}
\setlength{\tabcolsep}{1mm}{
\begin{tabular}{cccccc}
\hline
\toprule
 & Bona fide & Spoofed & Transition & Bona fide & Spoofed \\
\multirow{-2}{*}{Test Set} & Speech & Speech & Region & Non-speech & Non-speech \\\hline
\midrule
Dev. & \multicolumn{1}{r}{\cellcolor[HTML]{A8A8A8}12.22} & \multicolumn{1}{r}{\cellcolor[HTML]{999999}17.75} & \multicolumn{1}{r}{\cellcolor[HTML]{DBDBDB}-6.76} & \multicolumn{1}{r}{\cellcolor[HTML]{FFFFFF}-19.89} & \multicolumn{1}{r}{\cellcolor[HTML]{FFFFFF}-20.18} \\
Eval. & \multicolumn{1}{r}{\cellcolor[HTML]{9B9B9B}26.76} & \multicolumn{1}{r}{\cellcolor[HTML]{999999}27.66} & \multicolumn{1}{r}{\cellcolor[HTML]{CCCCCC}-10.42} & \multicolumn{1}{r}{\cellcolor[HTML]{FFFFFF}-49.19} & \multicolumn{1}{r}{\cellcolor[HTML]{E4E4E4}-28.16} \\
\bottomrule
\hline
\end{tabular}
}
\label{tab:wrong}
\vspace{-0.1 in}
\end{table}

We analyze partially spoofed samples where the PartialSpoof-trained SSL-Res1D model made incorrect and correct predictions, employing Grad-CAM with the spoof class as the target class.

Table~\ref{tab:wrong} shows RCQs when the model makes incorrect predictions.  It shows that the model's RCQ values for transition regions are much lower than for bona fide or spoofed speech. Comparing Table~\ref{tab:wrong} with Table~\ref{tab:quanti} of the same PartialSpoof-trained model, we notice that the model mainly pays attention to speech segments and neglects the transition regions when it classifies wrongly. This suggests that the model's inability to accurately focus on transition regions remarkably contributes to its mistaken identification of spoofed samples.

\begin{figure}[h]
\vspace{-0.05 in}
\centerline{\hspace{0.2cm}\includegraphics[scale=0.020]{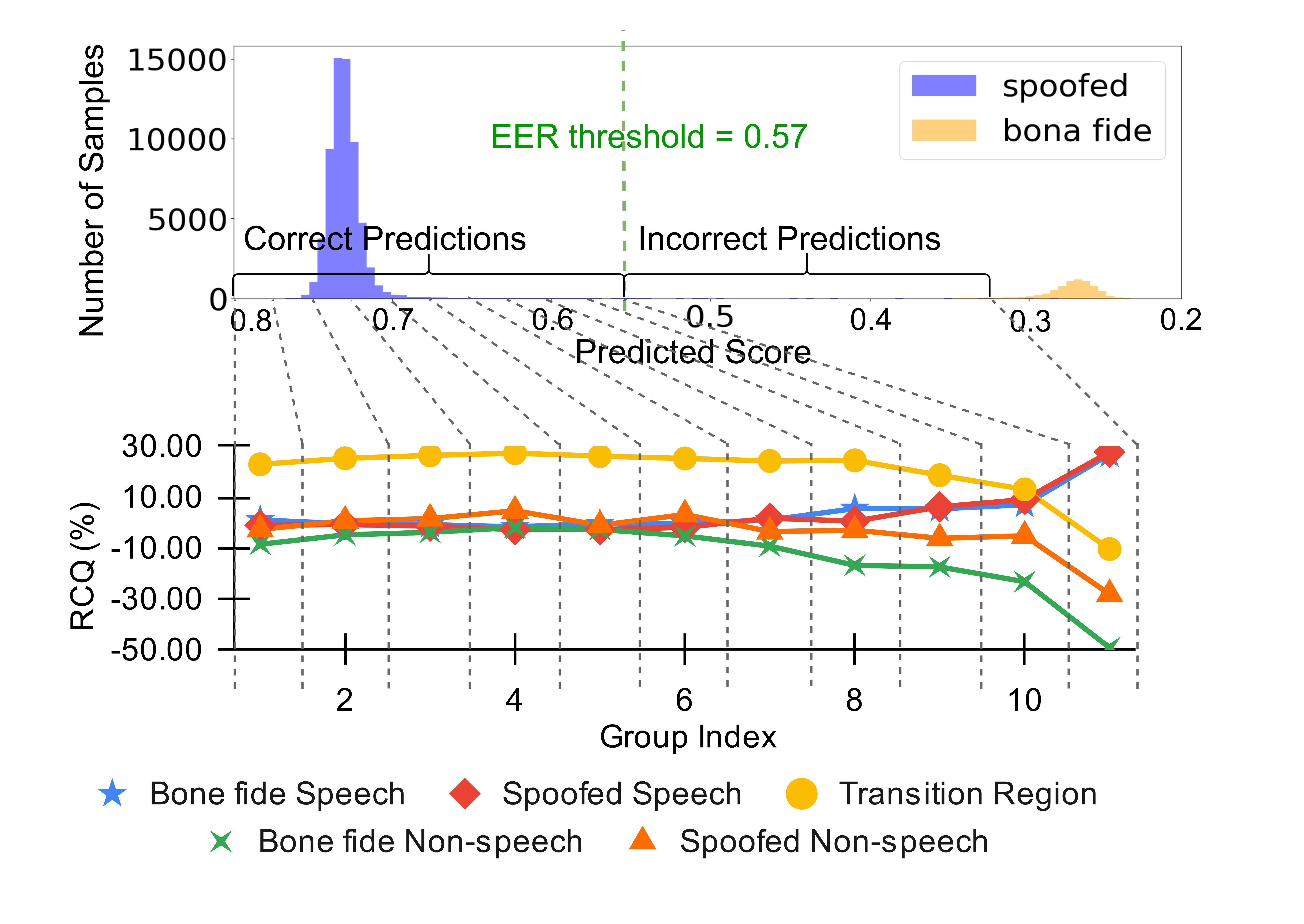}}
\vspace{-0.20 in}
\caption{The prediction score distribution of SSL-Res1D trained on PartialSpoof and tested on the evaluation set, along with the line chart showing the RQC of the five types of segments across 11 sample groups of partially spoofed samples.}
\label{fig_correct}
\vspace{-0.18 in}
\end{figure}

In addition to analyzing failure cases, we investigate how the CM's utterance-level prediction scores correlate with the focus areas in correctly predicted samples, as depicted in Figure~\ref{fig_correct}. In the upper panel of Figure~\ref{fig_correct}, we equally divide the range between the maximum and minimum predicted scores of the correctly predicted samples into ten groups.
The 11-th group consists of incorrectly classified samples. For each group, we analyze the focus areas of the CM using the RCQ, shown in the lower panel of Figure~\ref{fig_correct}. We notice that correctly classified cases in the model's predictions are linked to major attention on the transition regions. Conversely, as the output scores near the threshold, the model shifts its focus from the transition region to the speech parts. This trend highlights the importance of transition regions in the CM's performance, aligning with the aforementioned observation in Section~\ref{subsec_quantianaly}. 


\section{Conclusion}
In this study, we use Grad-CAM to explain how neural CMs detect partially spoofed audio. We observe that the partially spoofed data-trained CMs not only focus on the spoofed and bona fide areas, but also significantly on the transition regions created by the overlap-add operation during the dataset creation. We introduce a quantitative analysis metric, namely RCQ, to validate this observation across the entire evaluation set. Moreover, we conduct an analysis comparing correctly and incorrectly classified samples. We find that the CM pays less attention to the transition regions when it classifies incorrectly. In contrast, for the samples that are classified correctly, there is a higher focus on the transition regions for samples away from the EER threshold, confirming the important role of these transition regions in the effectiveness of the partial spoof CMs. These insights offer valuable contributions to the development of partial spoofing CMs and the creation of related datasets.



\balance
\bibliographystyle{IEEEtran}
\bibliography{mybib}

\begin{thebibliography}{10}
\providecommand{\url}[1]{#1}
\csname url@samestyle\endcsname
\providecommand{\newblock}{\relax}
\providecommand{\bibinfo}[2]{#2}
\providecommand{\BIBentrySTDinterwordspacing}{\spaceskip=0pt\relax}
\providecommand{\BIBentryALTinterwordstretchfactor}{4}
\providecommand{\BIBentryALTinterwordspacing}{\spaceskip=\fontdimen2\font plus
\BIBentryALTinterwordstretchfactor\fontdimen3\font minus \fontdimen4\font\relax}
\providecommand{\BIBforeignlanguage}[2]{{%
\expandafter\ifx\csname l@#1\endcsname\relax
\typeout{** WARNING: IEEEtran.bst: No hyphenation pattern has been}%
\typeout{** loaded for the language `#1'. Using the pattern for}%
\typeout{** the default language instead.}%
\else
\language=\csname l@#1\endcsname
\fi
#2}}
\providecommand{\BIBdecl}{\relax}
\BIBdecl

\bibitem{WU2015130}
Z.~Wu, N.~Evans, T.~Kinnunen, J.~Y. andFederico Alegre, and H.~Li, ``Spoofing and countermeasures for speaker verification: A survey,'' \emph{Speech Communication}, vol.~66, pp. 130--153, 2015.

\bibitem{bib:Attacker_overview2020}
R.~K. Das, X.~Tian, T.~Kinnunen, and H.~Li, ``The attacker's perspective on automatic speaker verification: An overview,'' in \emph{Proc. Interspeech}, 2020, pp. 4213--4217.

\bibitem{wangxininterspeech}
X.~Wang and J.~Yamagishi, ``A comparative study on recent neural spoofing countermeasures for synthetic speech detection,'' in \emph{Proc. Interspeech}, 2021, pp. 4259--4263.

\bibitem{tak22_odyssey}
H.~Tak, M.~Todisco, X.~Wang, J.~weon Jung, J.~Yamagishi, and N.~Evans, ``{Automatic} speaker verification spoofing and deepfake detection using wav2vec 2.0 and data augmentation,'' in \emph{Proc. Odyssey}, 2022, pp. 112--119.

\bibitem{10155166}
X.~Liu, X.~Wang, M.~Sahidullah, J.~Patino, H.~Delgado, T.~Kinnunen, M.~Todisco, J.~Yamagishi, N.~Evans, A.~Nautsch, and K.~A. Lee, ``{ASVspoof 2021}: Towards spoofed and deepfake speech detection in the wild,'' \emph{IEEE/ACM Transactions on Audio, Speech, and Language Processing}, vol.~31, pp. 2507--2522, 2023.

\bibitem{9417604}
Y.~Zhang, F.~Jiang, and Z.~Duan, ``One-class learning towards synthetic voice spoofing detection,'' \emph{IEEE Signal Processing Letters}, vol.~28, pp. 937--941, 2021.

\bibitem{9747766}
J.-w. Jung, H.-S. Heo, H.~Tak, H.-j. Shim, J.~S. Chung, B.-J. Lee, H.-J. Yu, and N.~Evans, ``{AASIST}: Audio anti-spoofing using integrated spectro-temporal graph attention networks,'' in \emph{Proc. ICASSP}, 2022, pp. 6367--6371.

\bibitem{kawa23b_interspeech}
P.~Kawa, M.~Plata, M.~Czuba, P.~Szymański, and P.~Syga, ``Improved deepfake detection using whisper features,'' in \emph{Proc. Interspeech}, 2023, pp. 4009--4013.

\bibitem{zang23_interspeech}
Y.~Zang, Y.~Zhang, and Z.~Duan, ``{Phase perturbation improves channel robustness for speech spoofing countermeasures},'' in \emph{Proc. Interspeech}, 2023, pp. 3162--3166.

\bibitem{zhanglin_PartialSpoof_TASLP}
L.~Zhang, X.~Wang, E.~Cooper, N.~Evans, and J.~Yamagishi, ``{The PartialSpoof} database and countermeasures for the detection of short fake speech segments embedded in an utterance,'' \emph{IEEE/ACM Transactions on Audio, Speech, and Language Processing}, vol.~31, pp. 813--825, 2023.

\bibitem{yi2023add}
J.~Yi \emph{et~al.}, ``{ADD 2023: The} second audio deepfake detection challenge,'' \emph{arXiv preprint arXiv:2305.13774}, 2023.

\bibitem{10094774}
Z.~Cai, W.~Wang, and M.~Li, ``Waveform boundary detection for partially spoofed audio,'' in \emph{Proc. ICASSP}, 2023, pp. 1--5.

\bibitem{muller2024mlaad}
N.~M. M{\"u}ller, P.~Kawa, W.~H. Choong, E.~Casanova, E.~G{\"o}lge, T.~M{\"u}ller, P.~Syga, P.~Sperl, and K.~B{\"o}ttinger, ``{MLAAD: T}he multi-language audio anti-spoofing dataset,'' \emph{arXiv preprint arXiv:2401.09512}, 2024.

\bibitem{Yi2021halftruth}
J.~Yi, Y.~Bai, J.~Tao, H.~Ma, Z.~Tian, C.~Wang, T.~Wang, and R.~Fu, ``{Half-Truth: A} partially fake audio detection dataset,'' in \emph{Proc. Interspeech}, 2021, pp. 1654--1658.

\bibitem{cai2023avdf1m}
Z.~Cai, S.~Ghosh, A.~P. Adatia, M.~Hayat, A.~Dhall, and K.~Stefanov, ``{AV-Deepfake1M}: A large-scale {LLM}-driven audio-visual deepfake dataset,'' \emph{arXiv preprint arXiv:2311.15308}, 2023.

\bibitem{Zhang2021PartialSpoofInitial}
L.~Zhang, X.~Wang, E.~Cooper, J.~Yamagishi, J.~Patino, and N.~Evans, ``An initial investigation for detecting partially spoofed audio,'' in \emph{Proc. Interspeech}, 2021, pp. 4264--4268.

\bibitem{martin2022ADD4}
J.~M. Martín-Doñas and A.~Álvarez, ``{The V}icomtech audio deepfake detection system based on wav2vec2 for the 2022 {ADD} challenge,'' in \emph{Proc. ICASSP}, 2022, pp. 9241--9245.

\bibitem{9746162}
H.~Wu, H.-C. Kuo, N.~Zheng, K.-H. Hung, H.-Y. Lee, Y.~Tsao, H.-M. Wang, and H.~Meng, ``Partially fake audio detection by self-attention-based fake span discovery,'' in \emph{Proc. ICASSP}, 2022, pp. 9236--9240.

\bibitem{CAI2024101597}
Z.~Cai and M.~Li, ``Integrating frame-level boundary detection and deepfake detection for locatingmanipulated regions in partially spoofed audio forgery attacks,'' \emph{Computer Speech \& Language}, vol.~85, p. 101597, 2024.

\bibitem{9746939}
J.~Yi \emph{et~al.}, ``{ADD} 2022: the first audio deep synthesis detection challenge,'' in \emph{Proc. ICASSP}, 2022, pp. 9216--9220.

\bibitem{silencegolden}
N.~Müller, F.~Dieckmann, P.~Czempin, R.~Canals, K.~Böttinger, and J.~Williams, ``{Speech is Silver, Silence is Golden: What do ASVspoof}-trained models really learn?'' in \emph{Proc. ASVspoof Challenge workshop}, 2021, pp. 55--60.

\bibitem{10224301}
Y.~Zhang, Z.~Li, J.~Lu, H.~Hua, W.~Wang, and P.~Zhang, ``The impact of silence on speech anti-spoofing,'' \emph{IEEE/ACM Transactions on Audio, Speech, and Language Processing}, vol.~31, pp. 3374--3389, 2023.

\bibitem{salvi2024listening}
D.~Salvi, T.~S. Balcha, P.~Bestagini, and S.~Tubaro, ``Listening between the lines: Synthetic speech detection disregarding verbal content,'' \emph{arXiv preprint arXiv:2402.05567}, 2024.

\bibitem{tak20_odyssey}
H.~Tak, J.~Patino, A.~Nautsch, N.~Evans, and M.~Todisco, ``An explainability study of the constant {Q} cepstral coefficient spoofing countermeasure for automatic speaker verification,'' in \emph{Proc. Odyssey}, 2020, pp. 333--340.

\bibitem{halpern20_odyssey}
B.~Halpern, F.~Kelly, R.~{van Son}, and A.~Alexander, ``Residual networks for resisting noise: Analysis of an embeddings-based spoofing countermeasure,'' in \emph{Proc. Odyssey}, 2020, pp. 326--332.

\bibitem{9747476}
W.~Ge, J.~Patino, M.~Todisco, and N.~Evans, ``Explaining deep learning models for spoofing and deepfake detection with shapley additive explanations,'' in \emph{Proc. ICASSP}, 2022, pp. 6387--6391.

\bibitem{ge22_odyssey}
W.~Ge, M.~Todisco, and N.~Evans, ``Explainable deepfake and spoofing detection: An attack analysis using {SHapley Additive exPlanations},'' in \emph{Proc. Odyssey}, 2022, pp. 70--76.

\bibitem{gradcam}
R.~R. Selvaraju, M.~Cogswell, A.~Das, R.~Vedantam, D.~Parikh, and D.~Batra, ``{Grad-CAM: V}isual explanations from deep networks via gradient-based localization,'' in \emph{Proc. ICCV}, 2017, pp. 618--626.

\bibitem{li2024phonemes}
P.~Li, T.~Wang, L.~Li, A.~Hamdulla, and D.~Wang, ``How phonemes contribute to deep speaker models?'' \emph{arXiv preprint arXiv:2402.02730}, 2024.

\bibitem{ma2024gradient}
Y.~Ma, K.~A. Lee, V.~Hautam{\"a}ki, M.~Ge, and H.~Li, ``Gradient weighting for speaker verification in extremely low signal-to-noise ratio,'' \emph{Proc. ICASSP}, 2024.

\bibitem{jacobgilpytorchcam}
J.~Gildenblat and contributors, ``Pytorch library for cam methods,'' \url{https://github.com/jacobgil/pytorch-grad-cam}, 2021.

\bibitem{NEURIPS2020_92d1e1eb}
A.~Baevski, Y.~Zhou, A.~Mohamed, and M.~Auli, ``wav2vec 2.0: A framework for self-supervised learning of speech representations,'' in \emph{Advances in Neural Information Processing Systems (NeurIPS)}, vol.~33, 2020, pp. 12\,449--12\,460.

\bibitem{NEURIPS2021_4cc05b35}
H.~Liu, Z.~Dai, D.~So, and Q.~V. Le, ``Pay attention to {MLPs},'' in \emph{Advances in Neural Information Processing Systems (NeurIPS)}, vol.~34, 2021, pp. 9204--9215.

\bibitem{10497864}
T.~Liu, K.~A. Lee, Q.~Wang, and H.~Li, ``Golden gemini is all you need: Finding the sweet spots for speaker verification,'' \emph{IEEE/ACM Transactions on Audio, Speech, and Language Processing}, vol.~32, pp. 2324--2337, 2024.

\bibitem{ECAPA_TDNN}
B.~Desplanques, J.~Thienpondt, and K.~Demuynck, ``{ECAPA-TDNN}: Emphasized channel attention, propagation and aggregation in {TDNN} based speaker verification,'' in \emph{Proc. Interspeech}, 2020, pp. 3830--3834.

\bibitem{MFA_TDNN}
T.~Liu, R.~K. Das, K.~A. Lee, and H.~Li, ``{MFA}: {TDNN} with multi-scale frequency-channel attention for text-independent speaker verification with short utterances,'' in \emph{Proc. ICASSP}, 2022, pp. 7517--7521.

\bibitem{liu2024disentangling}
T.~Liu, K.~A. Lee, Q.~Wang, and H.~Li, ``Disentangling voice and content with self-supervision for speaker recognition,'' in \emph{Advances in Neural Information Processing Systems (NeurIPS)}, vol.~36, 2023, pp. 50\,221--50\,236.

\bibitem{ASVspoof2019data}
X.~Wang \emph{et~al.}, ``{ASVspoof} 2019: A large-scale public database of synthesized, converted and replayed speech,'' \emph{Computer Speech and Language}, vol.~64, p. 101114, 2020.

\end{thebibliography}

\end{document}